\documentclass[prb,twocolumn,showpacs,amsmath,amssymb,superscriptaddress]{revtex4-2}

\usepackage{amsmath,amssymb,amsfonts,bm,color,graphicx,tabularx}

\usepackage[unicode=true,colorlinks=true]{hyperref}

\usepackage[etex=true,export]{adjustbox}
\usepackage{makecell}
\usepackage{multirow}
\usepackage{amsmath}
\usepackage{verbatim}

\hypersetup{linkcolor=blue, citecolor=blue,urlcolor=blue}
\usepackage{tabularx,multirow,array,diagbox}
\usepackage{adjustbox}

\newcommand{\beq}{\begin{equation}}
\newcommand{\eeq}{\end{equation}}
\newcommand{\beql}{\begin{equation*}}
\newcommand{\eeql}{\end{equation*}}
\newcommand{\beqn}{\begin{eqnarray}}
\newcommand{\eeqn}{\end{eqnarray}}

\begin{document}
\title{Majorana vortex phases in time-reversal invariant higher-order topological insulators and topologically trivial insulators}

\author{Xun-Jiang Luo}
\email{xjluo@hmfl.ac.cn}
\affiliation{Anhui Province Key Laboratory of Low-Energy Quantum Materials and Devices, 
High Magnetic Field Laboratory, HFIPS, Chinese Academy of Sciences, Hefei, Anhui 
230031, China }
\affiliation{Department of Physics, Hong Kong University of Science and Technology, Clear Water Bay, 999077 Hong Kong, China}
\author{Mingliang Tian}
\email{tianml@hmfl.ac.cn}
\affiliation{Anhui Province Key Laboratory of Low-Energy Quantum Materials and Devices, 
High Magnetic Field Laboratory, HFIPS, Chinese Academy of Sciences, Hefei, Anhui 
230031, China }
\affiliation{School of Physics and Optoelectronic Engineering, Anhui University, Hefei, 230601, 
China}

\begin{abstract}

Majorana vortex phases have been extensively studied in topological materials with conventional superconducting pairing. Inspired by recent experimental progress in realizing time-reversal invariant higher-order topological insulators (THOTIs) and inducing superconducting proximity effects, we investigate Majorana vortex phases in these systems. We construct THOTIs as two copies of a topological insulator (TI) with time-reversal symmetry-preserving mass terms that anisotropically gap the surface states. We find that these mass terms have a negligible impact on the vortex phase transitions of double TIs when treated as perturbations, and no additional topological phase transitions are induced.
Consequently,  $\mathbb{Z}_2$-protected Majorana vortex end modes (MVEMs) emerge when the chemical potential lies between the critical chemical potentials $\mu_c^{(1)}$ and $\mu_c^{(2)}$ of the two TI vortex phase transitions. We demonstrate this behavior across multiple THOTI models, including rotational symmetry-protected THOTI, inversion symmetry-protected THOTI, rotational and inversion symmetries-protected THOTI bismuth, and extrinsic THOTI. Remarkably, MVEMs persist even when all surfaces are gapped with the same sign, rendering the system topologically trivial in both first- and second-order classifications. 
Our findings establish that MVEMs can be realized in time-reversal invariant systems with fully gapped surfaces, encompassing both topologically nontrivial and trivial insulators, thus significantly broadening the solid state material platforms for hosting Majorana vortex phases.

\end{abstract}
\maketitle

\section{Introduction}

Topological superconductors (TSCs) have emerged as a central topic in condensed matter physics due to their ability to host Majorana zero modes (MZMs) \cite{Hasan2010,Qi2011}. These quasiparticles, with their non-Abelian statistics, offer promising applications for topological quantum computing \cite{Nayak2008}.  Although definitive proof of MZMs
remains elusive, substantial progress has been made in both theoretical \cite{Fu2008,Sau2010,QiXiaoLiang2010,Pientka2013,StevanNadjPerge2014,Xu2016,Pientka2017,PanXiaoHong2024,LuoMZM2024,zhang2025double,2026arXiv260315147L} and experimental 
 \cite{VMourik2012, XuJinPeng2015,SunHaoHua2016,Zhang2018,Fornieri2019,Liu2024} realms over the past decade. A hallmark experimental signature of MZMs is the observation of zero-bias conductance peaks (ZBCPs) in tunneling spectroscopy measurements \cite{LawKT2009}. These ZBCPs have been reported within Abrikosov vortices across diverse systems, including proximity-induced superconducting topological insulator (TI) Bi$_2$Te$_3$/NbSe$_2$ \cite{XuJinPeng2015,SunHaoHua2016},  topological crystalline insulator (TCI) SnTe/Pb \cite{Liu2024}, and intrinsic superconductors FeTe$_x$Se$_{1-x}$ \cite{Wang2018b,ShiyuZhu2020,Kong2019}, (Li$_{0.84}$Fe$_{0.16}$)OHFeSe \cite{LiuQin2018}, 2M-WS$_2$ \cite{Yuan2019,FanXuemin2024}, and LiFeAs \cite{Kong2021,Liu2022,Li2022}. Central to understanding these observations is the presence of a gapless surface Dirac cone in the normal-state, as first proposed by Fu and Kane for superconducting TI \cite{Fu2008}. Spin-momentum locking in the surface states induces an effective $p$-wave superconducting pairing, which enables the formation of MZMs at vortex cores \citep{Ivanov2001}. Fu-Kane mechanism
 was also extended to Dirac and Weyl semimetals \cite{QinShengshan2019,YanZhongbo2020,ZhangYi2022}, which host gapless bulk Dirac and Weyl cones, respectively. Despite these advances, a key question remains: can Majorana vortex end modes (MVEMs) be realized in time-reversal invariant systems, which are compatible with superconducting pairing, in the absence of both bulk and surface gapless Dirac cones?

\begin{figure*}[t]
\centering
\includegraphics[width=7in]{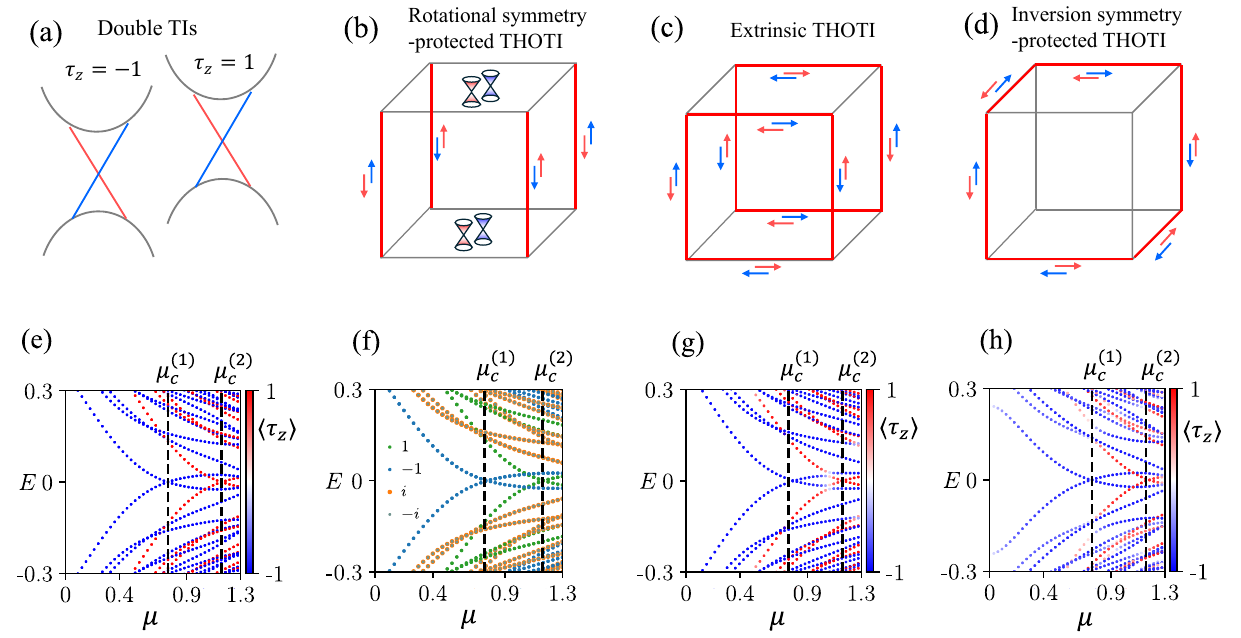}
\caption{ Schematical illustration for (a) double TIs residing in the subspace $\tau_z=1$ and $\tau_z=-1$, respectively, (b) rotational symmetry-protected THOTI, (c) extrinsic THOTI with trivial bulk topology, (d) inversion symmetry-protected THOTI. Red and blue arrows indicate hinge states with opposite flow directions.  (e)--(h) The energy spectrum of the BdG Hamiltonian \(\mathcal{H}_{\text{BdG}}(x, y, k_z = 0)\) as a function of chemical potential \(\mu\) associated with the systems depicted in (a)--(d), respectively. In (e), (g), and (h), the color represents the expectation value \(\langle \tau_z \rangle\). 
In (f), the color encodes the eigenvalue of the \(\mathcal{C}_{4z}\) symmetry. In our calculations, we take $\eta_0=0, \eta_1=0.4$ in (f),  $\eta_0=0.1,\eta_1=0.4$ in (g), and $V=0.1$ in (h). Common model parameters are \(\lambda = 0.2\), \(\Delta = 0.4\), and lattice size \(L_x = L_y = 30\). 
 }
\label{fig1}
\end{figure*}

Crystalline symmetries enrich the classification of topological phases, and their exploration leads to the discovery of higher-order topological insulators (HOTIs) \cite{Benalcazar2017a,Benalcazar2017,Langbehn2017,Fang2019a}. Over the past few years, higher-order topological phases, characterized by gapless modes at boundaries with a codimension greater than one, have attracted significant research interest due to their distinctive bulk-boundary correspondence \cite{Geier2018,Yan2018,Wang2018,Pan2019,Zhang2019,Zhang2019a,Xu2019,Yue2019,Sheng2019,Wu2020,chenli2021,Luo2021a,Pan2021,PhysRevB.106.L121108,Luo2023a,Luo2024,Jiazheng2024,PhysRevB.111.184516,2024BTIluo,zhang2024topological,zhang2024fermi,Luo2025}. A prominent class is time-reversal-invariant HOTIs (THOTIs) in three dimensions \cite{SongZhida2017,Schindler2018,Yoon20201}, which host gapped surfaces but feature robust helical hinge modes (HHMs). THOTIs can be realized in double band inversions systems with additional crystalline symmetries, such as rotation \cite{SongZhida2017,Fang2019a} or inversion \cite{Po2017,Khalaf2018,Khalaf2018a,Wang2019}. By using symmetry indicator theory and first-principles calculations,  THOTIs have been theoretically predicted in a diverse range of materials \cite{Tang2019,Tang2019a,Vergniory2019,CaoZhipeng2021}. Recently, significant experimental progress has been made in realizing THOTIs, with evidence of HHMs reported in  bismuth (Bi) \cite{Schindler2018a,Murani2019,Berthold2019,AbhayKumar2019,Aggarwal2021,Bernard2023}, Bi$_x$Sb$_{1-x}$ \cite{AbhayKumar2019}, 1T$^{\prime}$-XTe$_2$ (X=Mo,W) \cite{Kononov2020,Choi2020,Lee2023}, and Bi$_4$X$_4$ (X=Br, I)\cite{HuangJianwei2021,Noguchi2021,Shumiya2022,Zhao2023,Zhao2024,YuShuikang2024,Zhong2025}. In particular, Bi$_4$X$_4$ was predicted to exhibit superconductivity under pressure \cite{Qi2018,XiangLi2019}, and experiments have demonstrated superconducting proximity effects in Bi \cite{Drozdov2014,BertholdJäck2019,ChenXiaoyu2020,ChenKailun2024,LiuChen2024,2025arXiv250207533Z}.

\begin{figure}[t]
\centering
\includegraphics[width=3.2in]{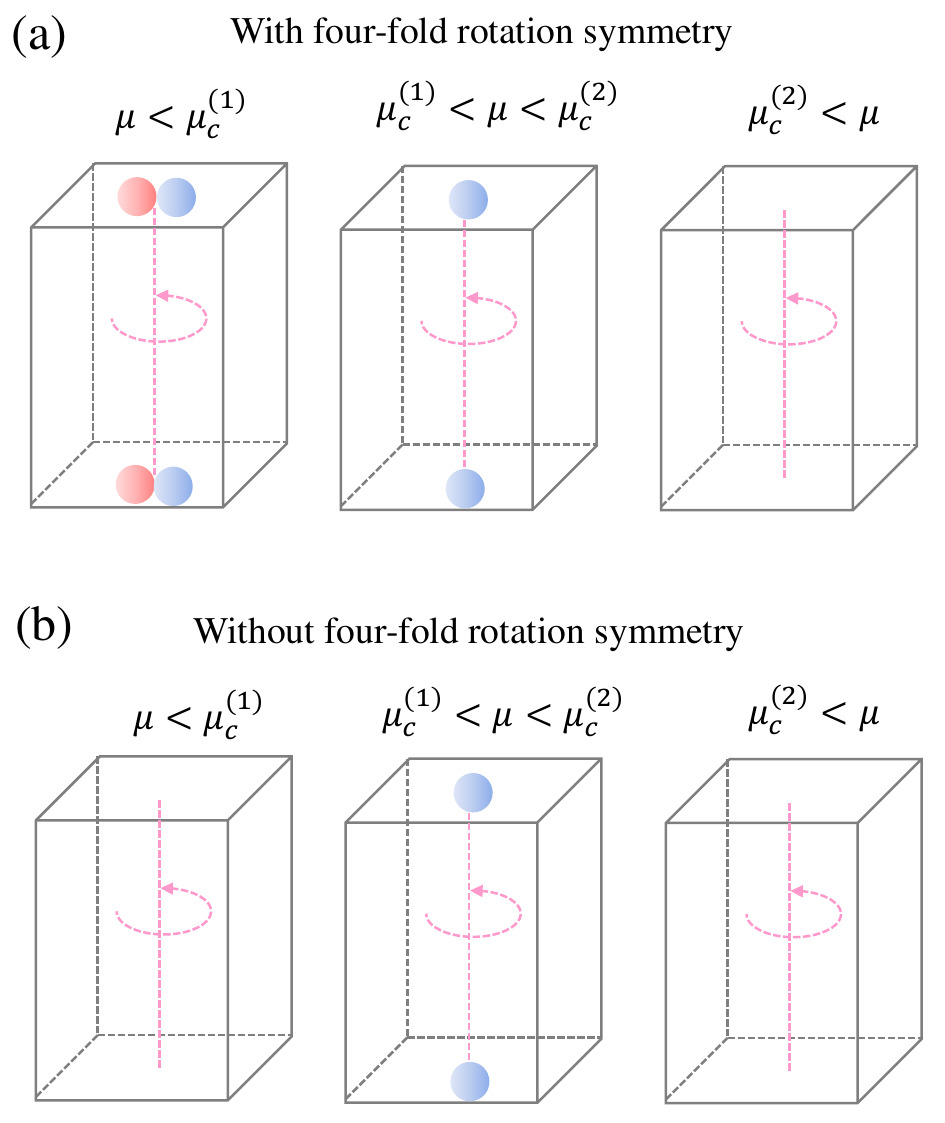}
\caption{(a) Schematical illustration of MVEMs for the case with the $\mathcal{C}_{4z}$ symmetry. The number of MVEMs is two for $\mu<\mu_c^{(1)}$,   one for $\mu_c^{(1)}<\mu<\mu_c^{(2)}$, and none for $\mu_c^{(2)}<\mu$. (b) Schematical illustration of MVEMs for the case without the $\mathcal{C}_{4z}$ symmetry and one MVEMs appear for $\mu_c^{(1)}<\mu<\mu_c^{(2)}$. }
\label{fig2}
\end{figure}

Motivated by recent experimental advances in superconducting THOTIs, we investigate Majorana vortex phases in these systems. We construct THOTIs by combining two decoupled TIs (double TIs)  [Fig.~\ref{fig1}(a)] and time-reversal symmetry-preserving mass terms that anisotropically gap the surface states, enabling the emergence of HHMs. This approach enables the realization of both intrinsic THOTIs \cite{Geier2018}, protected by rotational [Fig.~\ref{fig1}(b)] or inversion [Fig.~\ref{fig1}(d)] symmetry, and extrinsic THOTI [Fig.~\ref{fig1}(c)] with trivial bulk topology \cite{Geier2018}. In the absence of additional mass terms, the double TIs system undergoes two distinct vortex phase transitions at critical chemical potentials $\mu_c^{(1)}$ and $\mu_c^{(2)}$. Remarkably, the introduction of mass terms has a negligible impact on these transitions and does not induce additional topological phase transitions [Figs.~\ref{fig1}(e)-\ref{fig1}(h)]. Consequently, MVEMs emerge in THOTI systems when the chemical potential lies within the range $\mu_c^{(1)} < \mu < \mu_c^{(2)}$ [Fig.~\ref{fig2}(b)] due to the 
$Z_2$ classification for systems within the D symmetry class.  We further confirm this behavior in rotational and inversion symmetries-protected THOTI Bi \cite{Schindler2018a} through effective lattice model analysis, observing similar conditions for the emergence of MVEMs . Remarkably, MVEMs persist even when all surfaces of double TIs or TI are gapped with the same signs, rendering the system topologically trivial in both first- and second-order categories [Fig.~\ref{fig3}]. These models provide  paradigms for realizing Majorana vortex phases in topologically trivial systems.

This paper is organized as follows. In Sec.~\ref{secII}, we present a comprehensive construction of  THOTIs by introducing additional mass terms for double TIs. In Sec.~\ref{secIII}, we analyze the Majorana vortex phases in the THOTIs introduced in Sec.~\ref{secII}. In Sec.~\ref{secIV}, we study the Majorana vortex phase in superconducting THOTI Bi. In Sec.~\ref{secV}, we show that MVEMs can emerge in topologically trivial systems by uniformly gapping the surface states of double TIs or TI. In Sec.~\ref{secVI}, we provide a discussion and summary of our findings.  Appendices \ref{appendixa}-\ref{appendixc} complement the main text with additional technical details.

\section{Construction of THOTIs}
\label{secII}

 TIs are characterized by $\mathbb{Z}_2 \times \mathbb{Z}_2 \times \mathbb{Z}_2 \times \mathbb{Z}_2$ invariants, which can be determined by inversion eigenvalues at time-reversal invariant momenta if inversion symmetry is preserved \cite{Fu2007}. However, symmetry indicator theory predicts that time-reversal invariant systems possessing  inversion symmetry are described by a $\mathbb{Z}_2 \times \mathbb{Z}_2 \times \mathbb{Z}_2 \times \mathbb{Z}_4$ invariant structure \cite{Tang2019}. This discrepancy reveals a THOTI phase characterized by a $\mathbb{Z}_4$ symmetry indicator invariant $\kappa = 2$, arising from double band inversions at time-reversal invariant points.
Additionally, rotational symmetry-protected TCIs are predicted to host gapless Dirac cones on surfaces perpendicular to an $n$-fold rotational axis, which are connected by $n$ one-dimensional (1D) HHMs along hinges parallel to the rotation axis \cite{Fang2019a}. Consequently, such TCIs are also classified as THOTIs due to the presence of HHMs. Using a combination of symmetry indicator analysis and first-principles calculations, THOTIs protected by either inversion or rotational symmetry have been predicted in a wide range of materials \cite{Tang2019,Tang2019a,Vergniory2019,CaoZhipeng2021}. In the following, we construct intrinsic THOTIs stabilized by these crystalline symmetries, as well as extrinsic THOTIs that lack crystalline symmetry protection.

We construct lattice models for THOTIs by coupling two copies of TIs. The Hamiltonians for rotational and inversion symmetry-protected THOTIs are, respectively,  defined as follows:
\beqn
&&H_{\mathrm{HOTI}}^{(1)}(\bm{k})=\tau_0\otimes H_{\text{TI}}(\bm k)+\eta(\bm k)\tau_y\sigma_ys_0+\lambda\tau_z\sigma_0s_0,\nonumber\\
&&H_{\mathrm{HOTI}}^{(2)}(\bm{k})=\tau_0\otimes H_{\text{TI}}(\bm k)+V\sum_{i=x,y,z}\tau_y\sigma_0s_i+\lambda\tau_z\sigma_0s_0,\nonumber\nonumber\\
&&H_{\text{TI}}(\bm k)=M(\bm k)\sigma_zs_0+\sum_{i=x,y,z}\sin k_i\sigma_xs_i,
\label{hahoti}
\eeqn
where the mass terms are given by
 $M(\bm{k}) = m_0 - \sum_{i=x,y,z} \cos k_i$ and $\eta(\bm{k}) =  \eta_0 +\eta_1(\cos k_x - \cos k_y)$. The Pauli matrices $\tau$, $\sigma$, and $s$ act on the subspaces of the two TI copies, the orbital, and the spin, respectively. $\eta_0$, $\eta_1$, and $V$ are model parameters and $\lambda$ is used to shift the energy bands of two TI copies. To ensure nontrivial band topology, we set $m_0 = 2.5$ throughout this paper. Under this setting,  $H_{\text{TI}}$ describes a standard TI with band inversion at the $\Gamma$ point ($\bm{k} = 0$).  
The  time-reversal symmetry is given by $T = i s_y K$, where $K$ denotes complex conjugation.

When $\eta_0 = 0$, the higher-order topology of $H_{\mathrm{HOTI}}^{(1)}$ was studied in Ref.~\onlinecite{SongZhida2017}. This THOTI is protected by the time-reversal symmetry and four-fold rotational symmetry around the $z$-axis, represented by $C_{4z} = \tau_z e^{-i \pi s_z / 4}$. In contrast, $H_{\mathrm{HOTI}}^{(2)}$ breaks the $C_{4z}$ symmetry but preserves the inversion symmetry, represented by $I=\tau_0\sigma_zs_0$. With $T$ and $I$ symmetries, $H_{\mathrm{HOTI}}^{(2)}$ has a $\mathbb{Z}_4$ classification and $\kappa=2$ for this symmetry indicator invariant because of the double band inversions at the $\Gamma$ point. This indicates that $H_{\mathrm{HOTI}}^{(2)}$ describes a inversion symmetry-protected THOTI \cite{Khalaf2018}.

The non-trivial higher-order topology of $H_{\mathrm{HOTI}}^{(1)}$  can be revealed from its surface state analysis. For $\eta(\bm{k}) = 0$, $H_{\mathrm{HOTI}}^{(1)}$ decouples into two copies of TIs. In this configuration, the system exhibits two gapless Dirac cones on each of the (100), (010), and (001) surfaces. Introducing a non-zero $\eta_1$ while maintaining $\eta_0 = 0$,  the two Dirac cones on the (100) and (010) surfaces are gapped with opposite mass signs, enforced by the $C_{4z}$ symmetry. This results in gapless HHMs along the $z$-direction. Since $\eta(\bm{k})=0$ at the $\Gamma$ point, the (001) surface remains gapless, hosting two Dirac cones.  A schematic illustration of this $C_{4z}$ symmetry-protected THOTI is presented in Fig.~\ref{fig1}(b). When $\eta_0$ is non-zero, the $C_{4z}$ symmetry is broken, leading to a gapped (001) surface with a gap magnitude of $\eta_0$. In this scenario, the two Dirac cones on the (100) and (010) surfaces acquire gaps of magnitudes $\eta_0 - \eta_1/2$ and $\eta_0 + \eta_1/2$, respectively (see Appendix~\ref{appendixa}). Consequently, for $0 < \eta_0 < \eta_1/2$, the (100) surface develops a gap with a sign opposite to that of the (010) and (001) surfaces. This configuration results in HHMs along the edges of the (100) surface, as schematically illustrated in Fig.~\ref{fig1}(c). 
Although the absence of crystalline symmetry protection, this extrinsic THOTI phase is protected by the time-reversal symmetry and surface energy gaps, ensuring the robustness of HHMs \cite{Geier2018}.

The higher-order topology of $H_{\mathrm{HOTI}}^{(2)}$ can be understood analogously. A finite $V$ induces Dirac masses with opposite signs on the inversion-related surfaces, enforced by the inversion symmetry, while maintaining the same mass at the surfaces $x_i = L$ (see  Appendix~\ref{appendixa}). Here, $x_i=x,y,z$ and $L$ denotes the system size along each direction. This surface energy gap configuration generates HHMs along specific direction, as schematically depicted in Fig.~\ref{fig1}(d). These HHMs are protected by both inversion and time-reversal symmetries.
Notably, the discussion of the higher-order topology assumes $\lambda = 0$, and a small $\lambda$ does not significantly alter the underlying topological properties. However,  a finite $\lambda$ is important for the emergence of $\mathbb{Z}_2$-protected MVEM, as demonstrated in Sec.~\ref{secIII}.

\begin{figure}[t]
\centering
\includegraphics[width=3.2in]{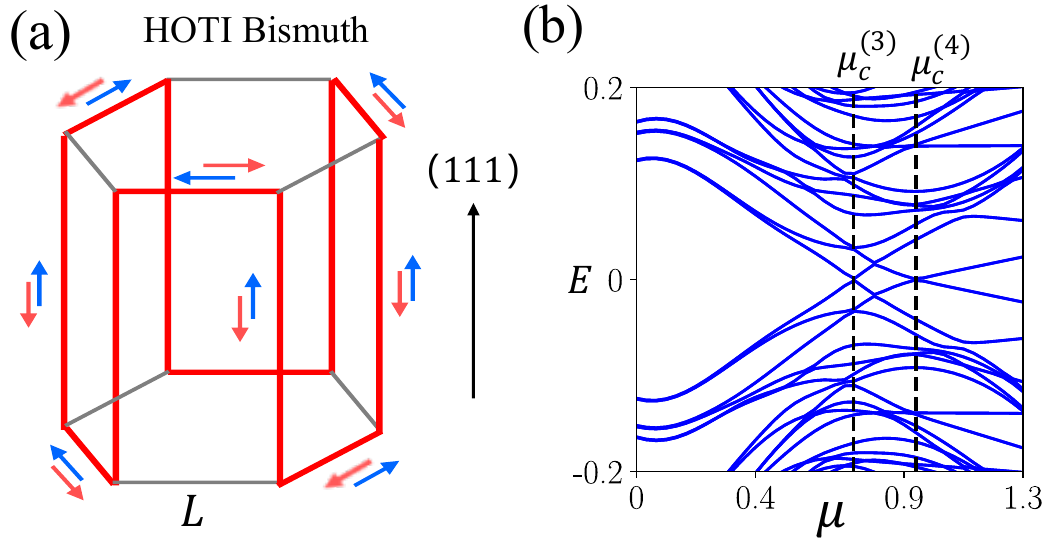}
\caption{(a) Schematical illustration of HHMs in THOTI Bi. $L$ denotes the  length of side of the (111) surface. (b) Vortex energy spectra of $\mathcal{H}_{\text{BdG}}(x,y,k_z=\pi)$, as a function of $\mu$, for superconducting bismuth. In (b),  $L=24$ and $\Delta=0.4$.}
\label{fig4}
\end{figure}

\section{Majorana vortex phases in THOTIs }
\label{secIII}

We now employ the THOTIs, illustrated in Figs.~\ref{fig1}(b)-\ref{fig1}(d), to explore Majorana vortex phases. In a 3D Bogoliubov-de Gennes (BdG) system, a superconducting vortex along the $z$-direction breaks the in-plane translational symmetry.  As a result, the entire system can be effectively treated as a 1D system. In the Nambu basis $\boldsymbol{\Psi} = (\boldsymbol{\psi}, i s_y \boldsymbol{\psi}^\dagger)^T$, with $\boldsymbol{\psi}$ being the basis of the normal-state Hamiltonian, the BdG Hamiltonian for this effective 1D system can be written as,
\begin{equation}
\mathcal{H}_{\mathrm{BdG}}(\bm{r}, k_z) = \begin{pmatrix}
H_{\mathrm{HOTI}}(\bm{r}, k_z) - \mu & \Delta(\bm{r}) \\
\Delta^\dagger(\bm{r}) & -H_{\mathrm{HOTI}}(\bm{r}, k_z) + \mu
\end{pmatrix},
\label{hbdg}
\end{equation}
where $\mu$ represents the chemical potential, and the superconducting pairing potential is expressed as $\Delta(\bm{r}) = \Delta \tanh\left({r}/{\xi}\right) e^{-i \theta}$. Here, $r = \sqrt{x^2 + y^2}$ is the radial distance in the $xy$-plane, $\theta = \arctan\left({y}/{x}\right)$ is the azimuthal angle, and $\xi$ denotes the superconducting coherence length, which we set to $\xi = 2$ without loss of generality. The particle-hole symmetry is given by $P=\rho_ys_yK$, where $\rho_y$ denotes the Pauli matrix acting on particle-hole subspace.
$\mathcal{H}_{\text{BdG}}$ belongs to the D symmetry class and has a $Z_2$ topological classification \cite{Chiu2016}.   In Eq.~\eqref{hbdg}, we neglect contributions from the vector potential and Zeeman term associated with the magnetic field inducing the vortex, as these are assumed to have minimal impact on the topological properties of interest.

We first analyze the Majorana vortex phases in superconducting THOTIs from surface states perspective. For the double TIs system depicted in Fig.~\ref{fig1}(a), the (001) surface hosts two gapless Dirac cones, which give rise to a pair of MZMs at a vortex core according to the Fu-Kane mechanism \cite{Fu2008}. We use $\gamma_1$ and $\gamma_2$ to denote the two MZMs. Notably, the four-fold rotational symmetry of the double TIs system is preserved under the vortex line, and is represented by the operator $\mathcal{C}_{4z} = e^{-i \pi \rho_z / 4} \otimes C_{4z}$. 
By deriving the analytical wave function of $\gamma_1$ and $\gamma_2$ from surface state Hamiltonian, it can be shown that (see Appendix \ref{appendixb})
\beqn
\mathcal{C}_4 \gamma_1 \mathcal{C}_4 = \gamma_1, \quad \mathcal{C}_4 \gamma_2 \mathcal{C}_4 = -\gamma_2.
\eeqn
Therefore, $\gamma_1$ and $\gamma_2$ are the eigenstates of $\mathcal{C}_{4z}$ symmetry with eigenvalue 1 and -1, respectively, which implies that they are protected by the $\mathcal{C}_4$ symmetry and robust against symmetry-preserving perturbations \cite{Kobayashi2020,LuoXunJiang2025}. On the basis of double TIs, THOTIs depicted in Figs.~\ref{fig1}(c)-\ref{fig1}(d), are generated by considering additional mass terms. As the mass term in THOTI depicted in Fig.~\ref{fig1}(b) 
preserves the $C_{4z}$ symmetry, the two MZMs $\gamma_1$ and $\gamma_2$ persist in this system, which is consistent with the existence of two gapless Dirac cones for normal-state. In contrast, the mass terms presented in THOTIs depicted in Figs.~\ref{fig1}(c) and \ref{fig1}(d) break the $C_{4z}$ symmetry, and therefore $\gamma_1$ and $\gamma_2$ are coupled by these additional mass terms and finally no MVEM persists. It is noted that this analysis describes the low-energy physics when the chemical potential $\mu$ crosses the surface states. However, it does not fully capture the vortex phase transitions that occur as $\mu$ increases.

\begin{figure*}[t]
\centering
\includegraphics[width=7in]{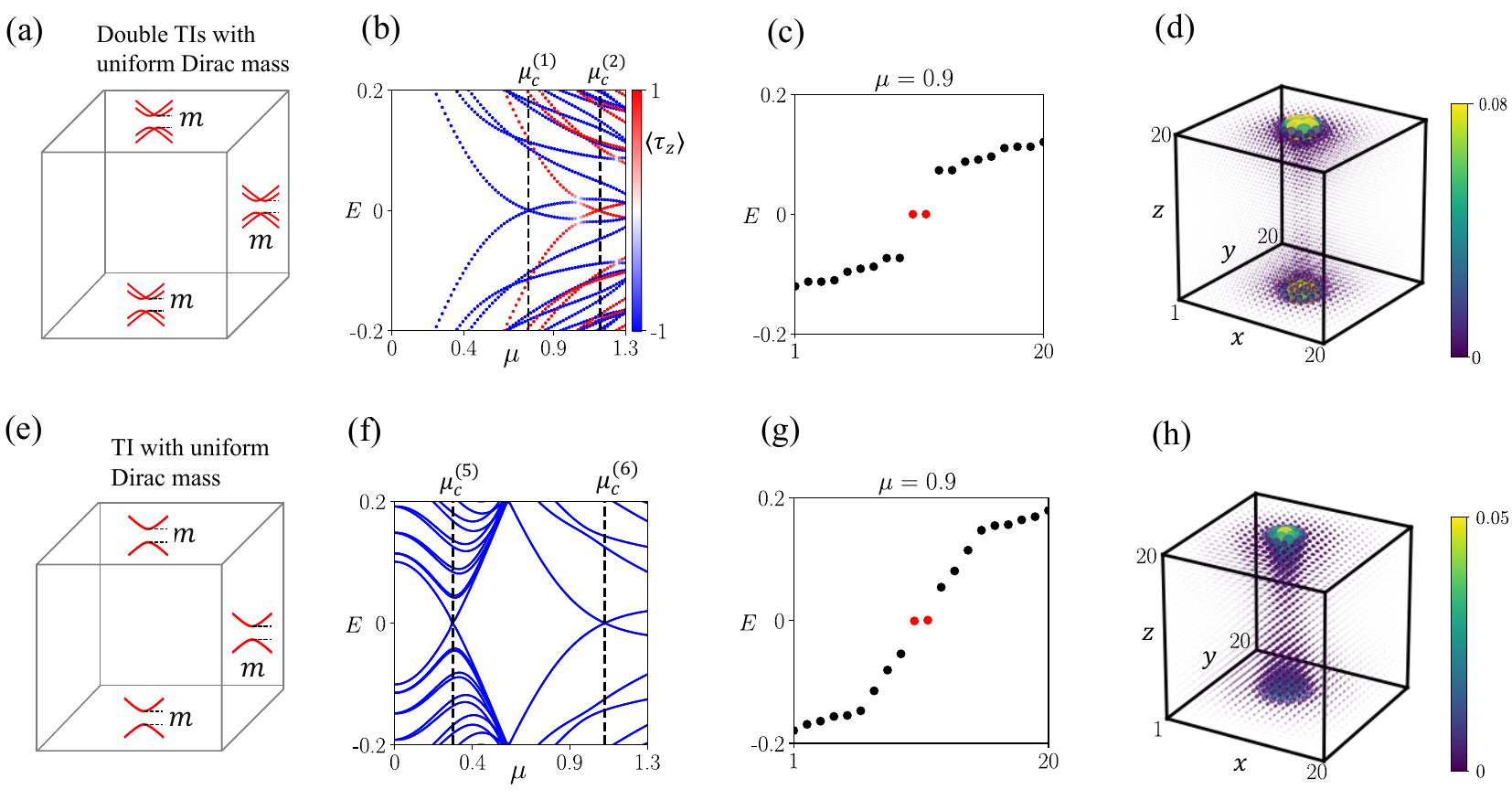}
\caption{ (a) and(e) Schematical illustration of trivial insulators for double TIs and TI, respectively, with uniform Dirac mass $m$ for surfaces states. (b) and (f) Vortex energy spectra of $\mathcal{H}_{\text{BdG}}(x,y,k_z=0)$ associated with the systems depicted in (a) and (e), respectively. (c) and (g) Energies close to zero for the superconducting trivial systems in (a) and (e), respectively, under the open boundary conditions. (d) and (h) The spatial wave function distribution of the two MVEMs, represented red dots in (c) and (g), respectively. In (b)-(d), $\eta_0=0.1$, $\eta_1=0$, $\lambda=0.2$, and $\Delta=0.4$. In (f)-(h), $m=0.5$ and $\Delta=0.4$. }
\label{fig3}
\end{figure*}

We now study the vortex phase transitions of superconducting THOTIs and compare them to those in double TIs. We numerically diagonalize the 3D BdG Hamiltonian $\mathcal{H}_{\text{BdG}}(x, y, k_z=0)$ using four distinct configurations of  $H_{\text{HOTI}}$:  (i) two copies of a TI with shifted energy bands ($\lambda\neq 0$), as illustrated in Fig.~\ref{fig1}(a); (ii)-(iv) the THOTI systems depicted in Figs.~\ref{fig1}(b)-\ref{fig1}(d), respectively. The vortex energy spectra as a function of $\mu$ for these four cases are shown in Figs.~\ref{fig1}(e)-\ref{fig1}(h), respectively.  In case (i), the shifted energy bands of two TI copies result in two distinct vortex phase transitions at $\mu_c^{(1)}$ and $\mu_c^{(2)}$, respectively. Remarkably, for cases (ii)-(iv), two vortex phase transitions occur near $\mu_c^{(1)}$ and $\mu_c^{(2)}$ with minimal modifications, and no additional vortex phase transitions are observed. This indicates that the additional mass terms $\eta(\bm{k})\tau_y\sigma_y s_0$ and $V\tau_y\sigma_0 s_{x,y,z}$ in Eq.~\eqref{hahoti} have a negligible impact on the vortex phase transitions of double TIs under the chosen model parameters, despite these terms drive the system to be a THOTI. This phenomenon can be qualitatively understood from perturbation theory. For the double TIs system illustrated in Fig.~\ref{fig1}(a), the two TI copies reside in the subspaces corresponding to $\tau_z = 1$ and $\tau_z = -1$, respectively. Upon introducing a vortex line, the vortex-bound states in this system are eigenstates of the $\tau_z$ operator, as indicated by the color coding in Fig.~\ref{fig1}(e). Specifically, the two vortex phase transitions, occurring at $\mu_c^{(1)}$ and $\mu_c^{(2)}$, are distinguished by their respective $\tau_z$ eigenvalues. Consequently, the two degenerate zero-energy states with $\tau_z = -1$ at $\mu_c^{(1)}$ or with $\tau_z = 1$ at $\mu_c^{(2)}$ cannot be coupled by the additional mass terms $\eta(\bm{k})\tau_y\sigma_y s_0$ or $V\tau_y\sigma_0 s_{x,y,z}$,  leaving the vortex phase transitions unaffected.  It is noted that the above discussion is applicable only when the additional mass terms are perturbations. Otherwise, the values of critical chemical potentials of the two vortex phases transitions are modified or even the system are trivialized by the additional mass terms (see Appendix~\ref{appendixd} ).

Notably, although the superconducting THOTI systems in cases (ii)-(iv) exhibit identical vortex phase transitions, the TSC phases for case (ii) differ from those in cases (iii) and (iv). In case (ii), the system hosts the $\mathcal{C}_{4z}$ symmetry, which enriches the topological classification from $\mathbb{Z}_2$ to $\mathbb{Z}_2 \otimes \mathbb{Z}_2$ \cite{Kobayashi2020}. In cases (iii) and (iv), the system hosts the $\mathbb{Z}_2$ classification within the D symmetry class. Specifically, in case (ii), the two vortex phase transitions are distinguished by the $\mathcal{C}_{4z}$ symmetry eigenvalues, as indicated by the color coding in Fig.~\ref{fig1}(f). This implies that, for $\mu<\mu_c^{(1)}$, both the Hamiltonians, in the subsectors having  the $\mathcal{C}_{4z}$ symmetry eigenvalues $1$ and $-1$, are topologically nontrivial, and the two subsectors cannot be coupled because of the $\mathcal{C}_{4z}$ symmetry.
Consequently, the number of robust MVEMs changes as $2\rightarrow 1\rightarrow 0$ when $\mu$ crosses ${\mu}_c^{(1)}$ and ${\mu}_c^{(2)}$ in sequence,  as schematically depicted in Fig.~\ref{fig2}(a).
In contrast, for cases (iii) and (iv), no crystalline symmetries can protect two MVEMs. Consequently, MVEMs are present on the (001) surface only within the region ${\mu}_c^{(1)}<\mu<{\mu}_c^{(2)}$, as schematically illustrated in Fig.~\ref{fig2}(b). This scenario is also applicable to the 
side surfaces of systems in cases (ii)-(iv), owing to the absence of rotational symmetry for these surfaces.
It is noted that when $\lambda=0$, ${\mu}_c^{(1)}={\mu}_c^{(2)}$, and therefore no  $\mathbb{Z}_2$-protected MVEMs exist.

\section{MVEMs in superconducting bismuth}
\label{secIV}

Bismuth-based materials have played a pivotal role in the advancement of topological physics \cite{Xia2009,ZKLiu2014}. However, pure bismuth, exhibiting two band inversions \citep{TeoJeffrey2008}, was long been thought to
be topologically trivial due to the $Z_2$ classification for strong TI. The discovery of higher-order topological phases has reclassified bismuth as a THOTI, protected by inversion symmetry and three-fold rotational symmetry $C_{3z}$ along its trigonal axis \citep{Schindler2018a}. The higher-order topology of bismuth has been robustly validated through multiple experimental observations by the scanning tunneling microscopy measurements \cite{Schindler2018a,Berthold2019,AbhayKumar2019,Aggarwal2021,2025arXiv250207533Z} and Josephson junction studies \cite{Schindler2018a,Murani2019,Bernard2023}.
Recently, the superconducting proximity effect in bismuth was demonstrated \citep{2025arXiv250207533Z}, and ZBCP, indicative of MZMs, was observed at the interface between a superconductor and an Fe cluster at the edge of bismuth \citep{BertholdJäck2019}.
Especially, ZBCPs were observed for Bi islands on superconductors Fe(Te,Se) \cite{ChenXiaoyu2020,ChenKailun2024}.
In the following, we employ the effective tight-binding model developed in Ref.~\onlinecite{Schindler2018a}, which captures the higher-order topology of bismuth, to investigate the Majorana vortex phase.

The effective model Hamiltonian for capturing higher-order topology of bismuth is defined on a simple 3D hexagonal lattice and can be written as \cite{Schindler2018a}
\beqn
H_\mathrm{TB}(\bm{k}) =&\,
\begin{pmatrix}
H_\mathrm{TB, I}(\bm{k}) + \epsilon & \delta \, M_\mathrm{TB} (\bm{k}) \\ \delta \, M_\mathrm{TB} ^\dagger(\bm{k}) & H_\mathrm{TB, II}(\bm{k}) - \epsilon
\end{pmatrix},
\eeqn
where $H_\mathrm{TB, I}(\bm{k})$ and $H_\mathrm{TB, II}(\bm{k})$ describe a strong TI within distinct eigenvalue sectors of the $C_{3z}$ symmetry. Specifically, $H_\mathrm{TB, I}$ corresponds to $C_{3z}$ eigenvalues $\exp(\pm i \pi / 3)$, and $H_\mathrm{TB, II}$ corresponds to an eigenvalue of $-1$. The $C_{3z}$ symmetry operator is expressed as $C_{3z} = C_{3,\mathrm{I}}^{z} \oplus  C_{3,\mathrm{II}}^{z}$,  with $ C_{3,\mathrm{I}}^{z} = \sigma_0 \otimes e^{i \frac{\pi}{3} \sigma_3}$ and $ C_{3,\mathrm{II}}^{z} = -\sigma_0 \otimes \sigma_0$. The term $M_\mathrm{TB}(\bm{k})$ describes the coupling between $H_\mathrm{TB, I}$ and $H_\mathrm{TB, II}$, with the coupling strength controlled by the parameter $\delta$. The parameter $\epsilon$ introduces a relative energy shift between $H_\mathrm{TB, I}$ and $H_\mathrm{TB, II}$. 
 Detailed expressions for $H_\mathrm{TB, I}(\bm{k})$, $H_\mathrm{TB, II}(\bm{k})$, and $M_\mathrm{TB}(\bm{k})$, along with specific model parameters, are provided in Appendix~\ref{appendixc}. When $\epsilon = 0$, the surface states of bismuth feature two Dirac points, each contributed by $H_\mathrm{TB, I}$ and $H_\mathrm{TB, II}$, respectively, and pinned at the same energy. In this scenario, the two Dirac cones cannot be gapped while preserving both $C_{3z}$ symmetry (preventing inter-cone gapping) and time-reversal symmetry (preventing intra-cone gapping). However,  no symmetry constraints fix the energies of the two Dirac cones, allowing them to be gapped for $\epsilon \neq 0$ \cite{Schindler2018a}. Enforced by the inversion and $C_{3z}$ symmetries, the surface states related by inversion exhibit opposite mass signs, while those related by three-fold rotation share identical mass signs. This energy gap configuration of the surface states gives rise to HHMs along specific direction, as schematically illustrated in Fig.~\ref{fig4}(a).

By replacing \( H_{\text{HOTI}} \) with \( H_{\text{TB}} \) in Eq.~\eqref{hbdg}, we diagonalize the BdG Hamiltonian \(\mathcal{H}_{\text{BdG}}(x, y, k_z = \pi)\) and identify two vortex phase transitions at chemical potentials \(\mu_c^{(3)}\) and \(\mu_c^{(4)}\), respectively. These transitions arise at \( k_z = \pi \) due to double band inversions at the \( T \) point [\(\bm{k} = (0, 0, \pi)\)] under the selected model parameters \cite{Schindler2018a}. This result aligns with the analysis in Sec.~\ref{secIII}. Unlike the \( C_{4z} \) symmetry, the $C_{3z}$ symmetry in bismuth does not change the classification of MVEMs \cite{Kobayashi2020,LuoXunJiang2025}, hosting the \(\mathbb{Z}_2\) topological classification. Consequently, a single MZM emerges at the vortex core when the chemical potential lies within \(\mu_c^{(3)} < \mu < \mu_c^{(4)}\). For bismuth, we note that surface relaxation effects can substantially modify the surface electronic structure and may influence the band topology \cite{Koie2025}. Such effects introduce additional complexity for the experimental realization of our proposal.

\section{MVEMs in topologically trivial insulators}
\label{secV}

The exploration of vortex line topology has traditionally focused on topological materials, where Majorana vortex phases are expected to emerge due to their nontrivial band structures.  Notably, Majorana-carrying vortices are not exclusive to systems with nontrivial bulk topology; it has been demonstrated that they can also emerge in topologically trivial quantum materials, such as HgTe \citep{ChiuChingKai2012,Hu2023}. Building on our prior analysis of vortex phase transitions in superconducting THOTIs, we now investigate the Majorana vortex phases in topologically trivial insulators.

For the Hamiltonian $H_{\text{HOTI}}^{(1)}$, we consider a configuration with $\eta_1=0$ and $\eta_0=0.1$. In this case, all the surface Dirac cones are gapped with the same sign, as schematically illustrated in Fig.~\ref{fig3}(a). Consequently, the system lacks gapless boundary states and is topologically trivial in both first-order and second-order topological classifications. Despite its trivial topology, numerical diagonalization of the BdG Hamiltonian  $\mathcal{H}_{\text{BdG}}(x, y, k_z=0)$ reveals two vortex phase transitions near $\mu_c^{(1)}$ and $\mu_c^{(2)}$, as shown in Fig.~\ref{fig3}(b). This result aligns with prior analysis indicating that the mass term $\eta_0 \tau_y \sigma_y s_0$ has a minimal impact on these transitions. Due to the $\mathbb{Z}_2$ topological classification within the D symmetry class, MVEMs emerge within the chemical potential range $\mu_c^{(1)} < \mu < \mu_c^{(2)}$.
In Figs.~\ref{fig3}(c) and  \ref{fig3}(d), we  numerically demonstrate the existence of MVEMs and show their spatial distribution, confirming their localized nature at the vortex ends.

Similarly, we consider a system described by the Hamiltonian \( H = H_{\text{TI}} + m \sigma_y s_0 \). The time-reversal symmetry breaking term \( m \sigma_y s_0 \) gaps all surface Dirac cones of \( H_{\text{TI}} \) with the same sign, as depicted in Fig.~\ref{fig3}(e), rendering the system topologically trivial with no gapless boundary modes. Substituting \( H_{\text{HOTI}} \) with \( H \) in Eq.~\eqref{hbdg}, numerical diagonalization of \( \mathcal{H}_{\text{BdG}}(x, y, k_z=0) \) identifies two vortex phase transitions at \( \mu_c^{(5)} \) and \( \mu_c^{(6)} \), as shown in Fig.~\ref{fig3}(f). MVEMs emerge within the range \( \mu_c^{(5)} < \mu < \mu_c^{(6)} \), as demonstrated in Figs.~\ref{fig3}(g) and \ref{fig3}(h). It is noted that the origin of the phase transitions at \( \mu_c^{(5)} \) and \( \mu_c^{(6)} \) are different.  The transition at \( \mu_c^{(6)} \) occurs when \( \mu \) enters the bulk states, a typical behavior for TI \cite{Hosur2011}. In contrast, the transition at \( \mu_c^{(5)} \) results from the competition between the magnetic energy gap and superconducting pairing gap for surface states, which occurs at the condition \( (\mu_c^{(5)})^2 + \Delta^2 = m^2 \). Similar vortex phase transitions were also theoretically studied in magnetic HOTI \cite{Ghorashi2020} and FeTeSe with magnetic impurities \cite{WuXianxin2021}. Therefore, the higher-order topology is not necessary for the emergence of MVEMs and they can be realized in topologically trivial systems.
These findings show that MVEMs can be realized in the absence of gapless hinge states and significantly broaden the scope of vortex line topology beyond conventional topological materials.

\section{Discussion and Summary}
\label{secVI}
We emphasize that although we use THOTIs as examples to investigate the Majorana vortex phase, the physics explored in this work is broadly applicable to any system hosting double band inversions and time-reversal symmetry. Double TIs with only the time-reversal symmetry is topologically trivial. However, the presence of additional crystalline symmetries, such as rotation, inversion, mirror, rotoinversion, or nonsymmorphic symmetries, can give rise to stable topology, commonly referred to as TCIs \citep{Khalaf2018a}. In TCIs, surfaces that preserve these specific crystalline symmetries host gapless Dirac cones. The Fu-Kane mechanism supports the formation of MVEMs on these gapless surfaces. Our work further reveals that MVEMs can also be realized when surfaces become gapped due to the breaking of the corresponding crystalline symmetry. Remarkably, even when crystalline symmetry is broken at all boundaries, rendering the system topologically trivial, MVEMs can still persist. The robustness of MVEMs in such systems can be qualitatively analyzed as follows. In a double TIs system, two TI copies residing in subspaces labeled by $\tau_z = \pm 1$, additional mass terms that preserve the $T$ symmetry must take the form $\tau_y \otimes \Xi$, where $\{T,\Xi\}=0$. These mass terms couple the two TI subsystems. Consequently, at critical chemical potentials $\mu_c^{(1)}$ and $\mu_c^{(2)}$, corresponding to vortex phase transitions of the two copies of TIs, degenerate zero-energy states characterized by $\tau_z=-1$ or $\tau_z=1$ cannot be coupled, leaving the two vortex phase transitions. Moreover, since these time-reversal symmetry-preserving mass terms are compatible with superconducting pairing, no additional vortex phase transitions are induced. Thus, in systems with double band inversions and time-reversal symmetry, MVEMs can generally emerge when $\mu_{c}^{(1)}<\mu<\mu_{c}^{(2)}$. Given that over half of all known nonmagnetic materials are topological, either TIs or TCIs \citep{Wieder2022}, our study significantly broadens the range of material platforms available for realizing Majorana vortex phases in solid-state systems. Moreover, beyond time-reversal invariant materials, magnetic topological materials \citep{Ghorashi2020} and even magnetic topologically trivial insulators [Fig.~\ref{fig3}(e)] also provide a broad platform for studying Majorana vortex phases.

In our theoretical models, the presence of a single MZM at a vortex core requires that \(\mu_c^{(1)} \neq \mu_c^{(2)}\), a condition arising from \(\lambda \neq 0\). We emphasize that this requirement can generally be satisfied in solid-state materials exhibiting double band inversions. In an TI subsystem, the critical chemical potential of the vortex phase transition is determined by the properties of the bulk state, including the Fermi velocity and the bulk energy gap \cite{Hosur2011}. Within a system featuring double band inversions, these properties are generally different between the two TIs, naturally resulting in \(\mu_c^{(1)} \neq \mu_c^{(2)}\). In addition, the chemical potential $\mu$ must lies within the energy window bounded by $\mu_c^{(1)}$ and $\mu_c^{(2)}$. Although fulfilling this condition in realistic materials presents certain experimental challenges, it remains experimentally feasible. In superconducting thin flim, $\mu$ can be tuned using gate voltages, chemical doping, or the proximity effect from the substrate. The precise location of the topological region for $\mu$ depends on the specific band structures of materials; therefore, it would be interesting to compute the topological phase diagram for realistic material.

We emphasize that the emergence of MVEMs in our THOTI proposals shares the same underlying mechanism as in the Fu-Kane model: the spin-momentum locking of surface states gives rise to an effective $p$-wave pairing. In THOTIs, although gapless surface states are absent, the surfaces host gapped Dirac cones that retain spin-momentum locking. When proximity-coupled to an $s$-wave superconductor, these gapped surface states acquire an effective $p$-wave pairing, thereby enabling MVEMs. From a bulk perspective, a vortex line can be regarded as a quasi-one-dimensional superconducting system in the D symmetry class, which hosts a $\mathbb{Z}_2$ topological classification. When the bulk energy gap can be tuned, for instance, by varying the chemical potential, a topological phase transition can occur, leading to the appearance of MVEMs.

In summary, we present a comprehensive study of Majorana vortex phases in both intrinsic and extrinsic THOTIs, which can be viewed as double TIs with additional mass terms. We demonstrate that these additional mass terms induce only minor modifications to the vortex phase transitions of the two TI copies, allowing MVEMs to emerge under appropriate conditions. Furthermore, we show that Majorana vortex phases can be realized in topologically trivial insulators, including double TIs or TI with a uniform Dirac mass for surface states. 
Our findings substantially expand the scope of materials to realize Majorana vortex phases, encompassing both topologically nontrivial and trivial systems with gapped surfaces.

\section{Acknowledgments}
Xun-Jiang Luo thanks Fengcheng Wu,
K. T. Law, and Zhongyi Zhang for helpful discussions. Mingliang Tian acknowledges the support of National Key R\&D Program of the MOST of China (Grant
Nos. 2022YFA1602603).

\appendix
\section{Surface states projection analysis}
\label{appendixa}

Surface state projection analysis is a useful tool for understanding HOTIs from the perspective of boundary domain walls. In this Appendix, we perform a surface state projection analysis for the THOTI models depicted in Figs.~\ref{fig1}(b)-\ref{fig1}(d). The model Hamiltonians for the THOTIs are expressed as:
\begin{align}
H_{\mathrm{HOTI}}^{(1)}(\bm{k}) &= \tau_0 \otimes H_{\text{TI}}(\bm{k}) + \eta(\bm{k}) \tau_y \sigma_y s_0 + \lambda \tau_z \sigma_0 s_0, \nonumber \\
H_{\mathrm{HOTI}}^{(2)}(\bm{k}) &= \tau_0 \otimes H_{\text{TI}}(\bm{k}) + V \sum_{i=x,y,z} \tau_y \sigma_0 s_i + \lambda \tau_z \sigma_0 s_0, \nonumber \\
H_{\text{TI}}(\bm{k}) &= M(\bm{k}) \sigma_z s_0 + \sum_{i=x,y,z} \sin k_i \sigma_x s_i, \label{shahoti}
\end{align}
where the mass term is given by $M(\bm{k}) = m_0 - \sum_{i=x,y,z} \cos k_i$ and the coupling term is $\eta(\bm{k}) = \eta_0 + \eta_1 (\cos k_x - \cos k_y)$. The Pauli matrices $\tau$, $\sigma$, and $s$ operate on the subspaces of the two TI copies, the orbital, and the spin, respectively.

For $1 < m_0 < 3$, the term $\tau_0 \otimes H_{\text{TI}}$ exhibits a double band inversions at the $\Gamma$ point, resulting in two gapless surface Dirac cones. These surface states can be gapped by additional mass terms. For simplicity, we set $\lambda = 0$ in the following analysis and consider the (100) surface as an example. By expanding the Bloch Hamiltonian around the $\Gamma$ point and imposing open boundary conditions along the $x$-direction, we obtain $H_{\mathrm{HOTI}}^{(1)} = h_0 + h_p$, where:
\begin{align}
h_0 &= (\tilde{m} - \partial_x^2/2) \tau_0 \sigma_z s_0 - i \partial_x \tau_0 \sigma_x s_x, \nonumber \\
h_p &= (\eta_0 + \eta_1 \partial_x^2/2) \tau_y \sigma_y s_0 + k_y \tau_0 \sigma_x s_y + k_z \tau_0 \sigma_x s_z,
\end{align}
with $\tilde{m} = m_0 - 3$ and $\tilde{\eta} = \eta_0 + \eta_1$. The Hamiltonian $h_0$ hosts eight zero-energy states, which are eigenstates of the operator $\tau_0 \sigma_y s_x$. The wavefunctions of these zero-energy states can be analytically derived by solving the eigenvalue equation $h_0 \psi_{\alpha}(x) = 0$ \cite{Yan2018}. By projecting the perturbation Hamiltonian $h_p$ onto the subspace spanned by the zero-energy states localized at $x = L$, we obtain the Dirac mass for the surface states on the (100) surface, given by $\Delta_x = \eta_0 + \tilde{m} \eta_1$. Similarly, the Dirac masses for the (010) and (001) surfaces are $\Delta_y = \eta_0 - \tilde{m} \eta_1$ and $\Delta_z = \eta_0$, respectively. For $m_0 = 2.5$, these become $\Delta_x = \eta_0 - \eta_1/2$, $\Delta_y = \eta_0 + \eta_1/2$, and $\Delta_z = \eta_0$. Thus, for $0 < \eta_0 < \eta_1/2$, the (100) surface develops a gap with a sign opposite to that of the (010) and (001) surfaces, leading to HHMs, as illustrated in Fig.~\ref{fig1}(c).

In the case of $H_{\mathrm{HOTI}}^{(2)}$, the two surface Dirac cones are gapped by the mass terms $V \sum_{i=x,y,z} \tau_y \sigma_0 s_i$. We can employ a similar surface state projection analysis to derive the Dirac mass, but here we approach the problem from a symmetry perspective. The surface state Hamiltonian can be written as $\tilde{H} = \tilde{h} + \tilde{h}_m$, where $\tilde{h}$ describes the two gapless surface Dirac cones and $\tilde{h}_m$ represents the Dirac mass. The Hamiltonian $H_{\mathrm{HOTI}}^{(2)}$ respects inversion symmetry, represented by $I = \tau_0 \sigma_z s_0$, satisfying $I H_{\mathrm{HOTI}}^{(2)}(\bm{k}) I^{-1} = H_{\mathrm{HOTI}}^{(2)}(-\bm{k})$. Under inversion, the surface state Hamiltonian transforms as,
\begin{equation}
I (\tilde{h} + \tilde{h}_m) I^{-1} = -(\tilde{h} - \tilde{h}_m).
\end{equation}
Consequently, surfaces related by inversion have opposite Dirac masses. However, since the $x$, $y$, and $z$ directions in $H_{\mathrm{HOTI}}^{(2)}$ are equivalent, the surfaces at $x_i = L$ (for $x_i = x, y, z$) exhibit the same Dirac mass sign. As a result, $H_{\mathrm{HOTI}}^{(2)}$ hosts HHMs, as depicted in Fig.~\ref{fig1}(d).

\begin{figure}[t]
\centering
\includegraphics[width=3.4in]{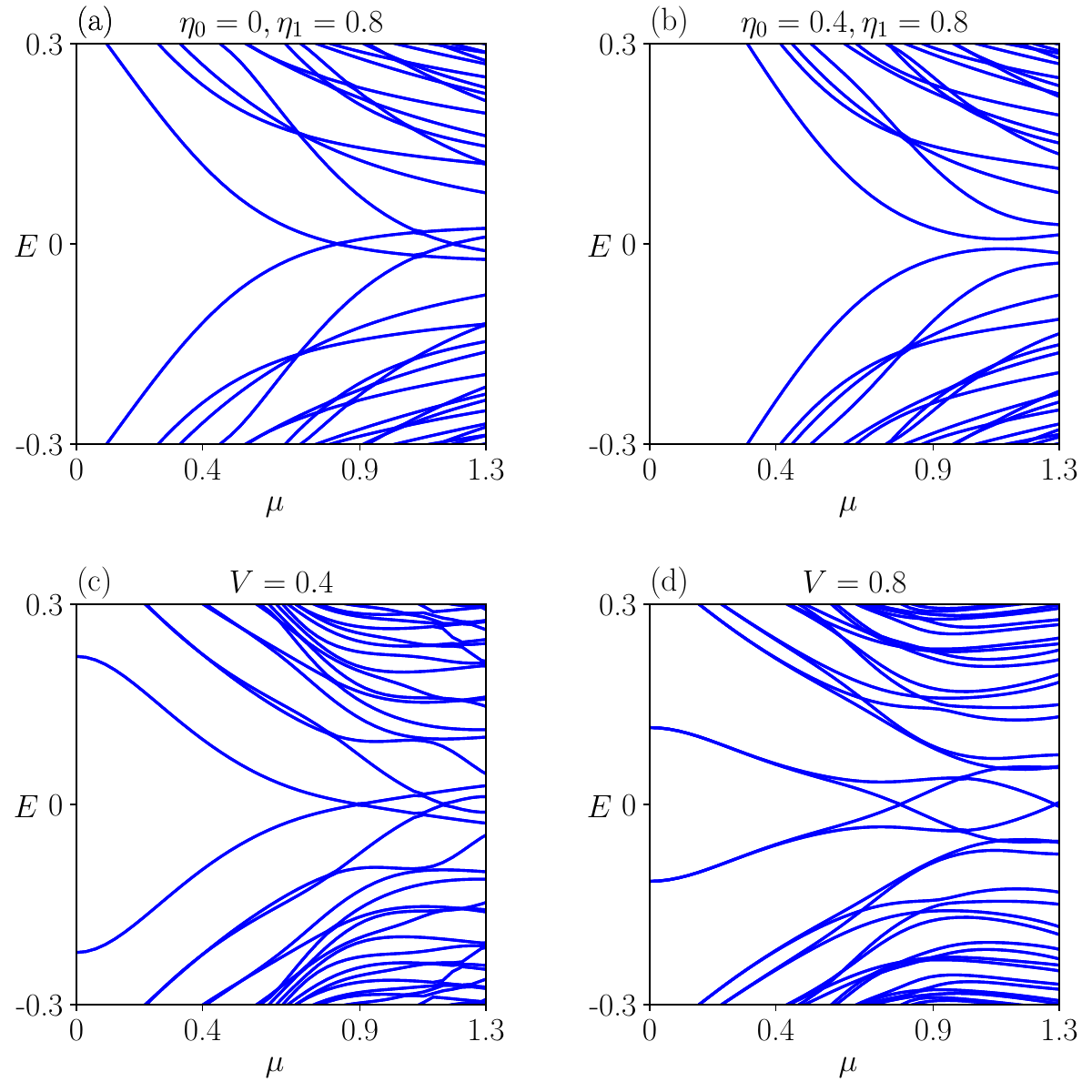}
\caption{(a) and (b) Vortex energy spectra of the Hamiltonian $\mathcal{H}_{\text{BdG}}(x,y,k_z=0)$ associated with the normal-state Hamiltonian $H_{\text{HOTI}}^{(1)}$. (c) and (d) Vortex energy spectra of the Hamiltonian $\mathcal{H}_{\text{BdG}}(x,y,k_z=0)$ associated with the normal-state Hamiltonian $H_{\text{HOTI}}^{(2)}$. The same model parameters are used as those in Fig.~\ref{fig1}.}
\label{fig5}
\end{figure}

\section{MZMs protected by rotational symmetry}

\label{appendixb}

For the double TIs system, the surface state Hamiltonian can be written as
\beqn
\tilde{H}=v_{F}k_x\tau_0s_x+v_{F}k_y\tau_0s_y,
\eeqn
where $v_{F}$ is the effective Fermi velocity.
When considering the superconducting vortex along the $z$-direction, the BdG Hamiltonian of describing surface states can be written as $H_{\text{BdG}}=\tau_0\otimes h_{\text{BdG}}$, where
\begin{equation}
\begin{aligned}
h_{\text{BdG}}&=-iv_F(\partial_x\rho_zs_x+\partial_y\rho_zs_y)\nonumber\\
&+\Delta_0\tanh(r/\xi)(\frac{x}{r}\rho_xs_0+\frac{y}{r}\rho_ys_0).
\end{aligned}
\end{equation}
 $H_{\text{BdG}}$ respects the particle-hole symmetry  $P=\tau_xK$ and four-fold rotational symmetry $\mathcal{C}_{4z} = e^{-i \pi \rho_z / 4}\tau_z e^{-i \pi s_z / 4}$. In the polar coordinates,  $H_{\text{BdG}}$ can be further written as
\begin{widetext}
\begin{equation}
\begin{aligned}
h_{\text{BdG}}=\left(\begin{array}{cccc}
0 & -iv_F e^{-i \theta}\left(\partial_r-\frac{i \partial_\theta}{r}\right) & \Delta(r) e^{-i \theta} & 0 \\
-iv_F e^{i \theta}\left(\partial_r+\frac{i \partial_\theta}{r}\right) & 0 & 0 & \Delta(r) e^{-i \theta} \\
\Delta(r) e^{i \theta}& 0 & 0 & iv_F e^{ -i\theta}\left(\partial_r-\frac{i \partial_\theta}{r}\right)\\
 & \Delta(r) e^{i \theta} & iv_F e^{i \theta}\left(\partial_r+\frac{i \partial_\theta}{r}\right) & 0
\end{array}\right),
\end{aligned}
\end{equation}    
\end{widetext}
where $h_{\text{BdG}}$ is written in the basis $\psi=\{c_{\uparrow},c_{\downarrow},-c_{\downarrow}^{\dagger}, c_{\uparrow}^{\dagger}\}$. Therefore, $h_{\text{BdG}}$ hosts the MZM solution
\beqn
\gamma=(e^{-i\pi/4}c_{\downarrow}+e^{i\pi/4}c_{\downarrow}^{\dagger})e^{-\int_{0}^{r}\Delta( r^{\prime})/v_Fdr^{\prime}},
\eeqn
where we have assumed that $v_{F}\Delta_0>0$. Since $H_{\text{BdG}}$ is just two copies of $h_{\text{BdG}}$, $H_{\text{BdG}}$ hosts two MZMs, denoted by $\gamma_1$ and $\gamma_2$ with the identical wave function in the particle-hole and spin space. While in the subspace $\tau_z=1$ and $\tau_z=-1$, the $C_{4z}$ symmetry is represented by $\mathcal{C}_{4z}=e^{-i \pi \rho_z}e^{-i \pi s_z / 4}$ and $\mathcal{C}_{4z}=-e^{-i \pi \rho_z}e^{-i \pi s_z / 4}$, respectively. It can be checked that 
\beqn
C_{4z}\gamma_1C_{4z}^{-1}=\gamma_1, \quad C_{4z}\gamma_2C_{4z}^{-1}=-\gamma_2,
\eeqn
which implies that MZMs $\gamma_1$ and $\gamma_2$ are protected by the $\mathcal{C}_{4z}$ symmetry.

\section{Effects of mass terms on vortex phase transitions }
\label{appendixd}
In the main text, we demonstrated that the additional mass terms \(\eta(\mathbf{k})\tau_y \sigma_y s_0\) and \(V \tau_y \sigma_0 s_{x,y,z}\) have a negligible impact on the vortex phase transitions of double TIs when treated as perturbations. In this appendix, we focus on the non-perturbative case. Based on double TIs, we consider the mass term \(0.8 (\cos k_x - \cos k_y) \rho_z \tau_y \sigma_y s_0\) and two vortex phases transition are observed with modified values of critical chemical potentials, as shown in Fig.~\ref{fig5}(a).
However, when considering the mass term \((0.4 + 0.8 (\cos k_x - \cos k_y)) \rho_z \tau_y \sigma_y s_0\), no vortex phase transition occurs, and the system becomes topologically trivial. This implies that the vortex phase transitions of double TIs are more sensitive to the \(\eta_0\)-dependent term than to the \(\eta_1\)-dependent term. This is because the vortex bound states are formed by the bulk states near the \(\Gamma\) point, and at this point, the \(\eta_1\)-dependent term vanishes. Similarly, we observe that the vortex phase transitions of double TIs are significantly modified by the mass terms \(0.4 \rho \tau_y \sigma_0 s_{x,y,z}\) and \(0.8 \rho \tau_y \sigma_0 s_{x,y,z}\), as shown in Fig.~\ref{fig5}(c) and Fig.~\ref{fig5}(d), respectively.

\section{Effective model Hamiltonian of HOTI Bi}
\label{appendixc}
The effective tight-binding Hamiltonian for describing the higher-order topology of Bismuth is defined on a 3D hexagonal lattice. The Hamiltonian is a $8 \times 8$ matrix in momentum space and is expressed as:
\begin{widetext}
\begin{equation}
\begin{aligned}
\label{eq: TBHamiltonian}
H_\mathrm{TB}(\bm{k}) =&\,
\begin{pmatrix}
H_\mathrm{TB, I}(\bm{k}) + \epsilon & \delta \, M_\mathrm{TB} (\bm{k}) \\ \delta \, M_\mathrm{TB} (\bm{k})^\dagger & H_\mathrm{TB, II}(\bm{k}) - \epsilon
\end{pmatrix}, \\
H_\mathrm{TB, I}(\bm{k}) =&\Gamma_1 \bigl\{m_\mathrm{I} (1 + \cos \bm{k}\cdot \bm{a}_3) - t_\mathrm{I} \left[\cos \bm{k}\cdot \bm{a}_1 + \cos \bm{k}\cdot \bm{a}_2 + \cos \bm{k}\cdot (\bm{a}_1 + \bm{a}_2) \right] \bigr\} \\&+ \lambda_\mathrm{I} \bigl[\Gamma_2 \sin \bm{k}\cdot \bm{a}_1 + \Gamma^{\mathrm{I},\mathrm{I}} _{2, 1} \sin \bm{k}\cdot \bm{a}_2 -  \Gamma^{\mathrm{I},\mathrm{I}} _{2, 2} \sin \bm{k}\cdot (\bm{a}_1 + \bm{a}_2) + \Gamma_3 \sin \bm{k}\cdot \bm{a}_3
\bigr],\\
H_\mathrm{TB, II}(\bm{k}) = &\Gamma_1 \bigl\{m_{\mathrm{II}} (1 + \cos \bm{k}\cdot \bm{a}_3) - t_{\mathrm{II}} \left[\cos \bm{k}\cdot \bm{a}_1 + \cos \bm{k}\cdot \bm{a}_2 + \cos \bm{k}\cdot (\bm{a}_1 + \bm{a}_2) \right] \bigr\}
\\&
+\lambda_{\mathrm{II}} \bigl[ \Gamma_2 \sin \bm{k}\cdot \bm{a}_1 + \Gamma^{\mathrm{II},\mathrm{II}} _{2, 1} \sin \bm{k}\cdot \bm{a}_2 - \Gamma^{\mathrm{II},\mathrm{II}} _{2, 2} \sin \bm{k}\cdot (\bm{a}_1 + \bm{a}_2) + \Gamma_3 \sin \bm{k}\cdot \bm{a}_3 \bigr]
\\ &
+ \Gamma_4 \gamma_{\mathrm{II}} \bigl[\sin \bm{k}\cdot (\bm{a}_1 + 2 \bm{a}_2) + \sin \bm{k}\cdot (\bm{a}_1 - \bm{a}_2) - \sin \bm{k}\cdot (2 \bm{a}_1 + \bm{a}_2) \bigr],
\\
M_\mathrm{TB}(\bm{k}) =&\, \Gamma_2 \bigl[\sin \bm{k} \cdot \bm{a}_1 + \sin \bm{k} \cdot (2\bm{a}_1 + \bm{a}_2)\bigr] + \Gamma^{\mathrm{I},\mathrm{II}} _{2, 1} \bigl[\sin \bm{k} \cdot \bm{a}_2 + \sin \bm{k} \cdot (\bm{a}_2 - \bm{a}_1)\bigr] 
\\&
- \Gamma^{\mathrm{I},\mathrm{II}} _{2, 2} \bigl[\sin \bm{k} \cdot (\bm{a}_1+\bm{a}_2) + \sin \bm{k} \cdot (\bm{a}_1 + 2\bm{a}_2)\bigr]
-\mathrm{i} \Gamma_5 \bigl[\cos \bm{k} \cdot \bm{a}_1 + \cos \bm{k} \cdot (2\bm{a}_1 + \bm{a}_2)\bigr] \\
&-\mathrm{i} \Gamma^{\mathrm{I},\mathrm{II}} _{5, 1} \bigl[\cos \bm{k} \cdot \bm{a}_2 + \cos \bm{k} \cdot (\bm{a}_2 - \bm{a}_1)\bigr] -\mathrm{i} \Gamma^{\mathrm{I},\mathrm{II}} _{5, 2} \bigl[\cos \bm{k} \cdot (\bm{a}_1+\bm{a}_2) + \cos \bm{k} \cdot (\bm{a}_1 + 2\bm{a}_2)\bigr].
\end{aligned}
\end{equation}
\end{widetext}
The Dirac matrices are defined as $\Gamma_1 = \sigma_3\otimes\sigma_0$, $\Gamma_2 = \sigma_1\otimes\sigma_1$, $\Gamma_3 = \sigma_2\otimes\sigma_0$, $\Gamma_4 = \sigma_1\otimes\sigma_2$, $\Gamma_5 = \sigma_3\otimes\sigma_1$, and
\begin{equation}
\Gamma_{\mu, \nu}^{i,\ j} = \left(C^{z}_{3,i} \right) ^{\nu} \Gamma_{\mu} \left(C^z_{3,j} \right)^{-\nu}.
\end{equation}
Here, $\mu \in \{ 1,\ \cdots , 5 \}$, $i, j \in \{\mathrm{I} ,\mathrm{II}  \}$, $\nu \in \{1,\ 2 \}$,  $C^z_{3,\mathrm{I}} = \sigma_0 \otimes e^{\mathrm{i} \frac{\pi}{3} \sigma_3}$ and $C^z_{3,\mathrm{II}} = - \sigma_0 \otimes \sigma_0$ so that the full threefold rotation symmetry is given by $C_{3z} = C^z_{3,\mathrm{I}} \oplus C^z_{3,\mathrm{II}}$. The lattice basis is $\bm{a}_1=(1,0,0)$, $\bm{a}_2=(-1/2,\sqrt{3}/2,0)$, and $\bm{a}_3=(0,0,1)$. Following Ref.~\cite{Schindler2018a}, we take $m_{\mathrm{I}}=m_{\mathrm{II}}=2, t_{\mathrm{I}}=t_{\mathrm{II}}=1, \lambda_{\mathrm{I}}=0.3, \lambda_{\mathrm{II}}=\gamma_{\mathrm{II}}=1, \epsilon=0.1 \text {, and } \delta=0.3$.

\bibliography{reference}

\end{document}